\documentclass[journal]{IEEEtai}
\usepackage[colorlinks,urlcolor=blue,linkcolor=blue,citecolor=blue]{hyperref}

\usepackage{color,array}
\usepackage{amsmath,algorithm,algorithmic}
\usepackage{float}
\usepackage{booktabs} % For prettier tables
\usepackage{graphicx}

\begin{document}

\title{A Brain-inspired Theory of Collective Mind Model for Efficient Social Cooperation}

\author{Zhuoya Zhao, Feifei Zhao, Shiwen Wang, Yinqian Sun and Yi Zeng
%\thanks{This study is supported by National Key Research and Development Program (Grant No. 2020AAA0107800), and the National Natural Science Foundation of China (Grant No. 62106261). }
\thanks{Zhuoya Zhao and Feifei Zhao contributed equally to this work.}
%\thanks{S. B. Author, Jr., was with Rice University, Houston, TX 77005 USA. He is now with the Department of Physics, Colorado State University, Fort Collins, CO 80523 USA (e-mail: author@lamar.colostate.edu).}
%\thanks{T. C. Author is with the Electrical Engineering Department, University of Colorado, Boulder, CO 80309 USA, on leave from the National Research Institute for Metals, Tsukuba, Japan (e-mail: author@nrim.go.jp).}
%\thanks{This paragraph will include the Associate Editor who handled your paper.}}
\thanks{Zhuoya Zhao is with Brain-inspired Cognitive Intelligence Lab, Institute of Automation, Chinese Academy of Sciences, Beijing 100190, China and School of Future Technology, University of Chinese Academy of Sciences, Beijing 100049, China. (e-mail:zhaozhuoya2019@ia.ac.cn)}
\thanks{Feifei Zhao is with Brain-inspired Cognitive Intelligence Lab, Institute of Automation, Chinese Academy of Sciences, Beijing 100190, China. (e-mail:zhaofeifei2014@ia.ac.cn)}
\thanks{Shiwen Wang is with School of Artificial Intelligence, University of Chinese Academy of Sciences, Beijing 100049, China. (e-mail:wangshiwen21@mails.ucas.ac.cn)}
\thanks{Yinqian Sun is with Brain-inspired Cognitive Intelligence Lab, Institute of Automation, Chinese Academy of Sciences, Beijing 100190, China. (e-mail:sunyinqian2018@ia.ac.cn)}
\thanks{Yi Zeng is with Brain-inspired Cognitive Intelligence Lab, Institute of Automation, Chinese Academy of Sciences, Beijing 100190, China, School of Future Technology, University of Chinese Academy of Sciences, Beijing 100049, China, School of Artificial Intelligence, University of Chinese Academy of Sciences, Beijing 100049, China, and Center for Excellence in Brain Science and Intelligence Technology, Chinese Academy of Sciences, Shanghai 200031, China.  (e-mail:yi.zeng@ia.ac.cn)}

\thanks{The corresponding author is Yi Zeng.}

}
%\thanks{This paragraph of the first footnote will contain the date on which you submitted your paper for review. It will also contain support information, including sponsor and financial support acknowledgment. For example, ``This work was supported in part by the U.S. Department of Commerce under Grant BS123456.'' }

%\markboth{Journal of IEEE Transactions on Artificial Intelligence, Vol. 00, No. 0, Month 2020}
%{Zhuoya Zhao \MakeLowercase{\textit{et al.}}: Brain-inspired ToCM}

\maketitle

\begin{abstract}
	Social intelligence manifests the capability, often referred to as the Theory of Mind (ToM), to discern others' behavioral intentions, beliefs, and other mental states. ToM is especially important in multi-agent and human-machine interaction environments because each agent needs to understand the mental states of other agents in order to better respond,  interact, and collaborate. Recent research indicates that the ToM model possesses the capability to infer beliefs, intentions, and anticipate future observations and actions; nonetheless, its deployment in tackling intricate tasks remains notably limited. The challenges arise when the number of agents increases, the environment becomes more complex, and interacting with the environment and predicting the mental state of each other becomes difficult and time consuming. To overcome such limits, we take inspiration from the Theory of Collective Mind (ToCM) mechanism, predicting observations of all other agents into a unified but plural representation and discerning how our own actions affect this mental state representation. Based on this foundation, we construct an imaginative space to simulate the multi-agent interaction process, thus improving the efficiency of cooperation among multiple agents in complex decision-making environments.
In various cooperative tasks with different numbers of agents, the experimental results highlight the superior cooperative efficiency and performance of our approach compared to the Multi-Agent Reinforcement Learning (MARL) baselines. We achieve consistent boost on SNN- and DNN-based decision networks, and demonstrate that ToCM's inferences about others' mental states  can be transferred to new tasks for quickly and flexible adaptation.
%There has been a burgeoning interest in multi-agent systems, which, aided by multi-agent reinforcement learning (MARL) algorithms, are finding expansive applications across various domains.
\end{abstract}

\begin{IEEEImpStatement}
This work delves into social intelligence through the lens of ToM and its application in multi-agent interactions, emphasizing its critical role in understanding and predicting the mental states of other entities. To address the limitations of ToM in complex tasks and environments, we introduce a brain-inspired ToCM model, which integrates the observations of all agents into a cohesive yet diverse mental state representation. This paper presents both the theoretical and practical models of ToCM which unifies the mental state representation of the collective, creating an imaginative space for efficient multi-agent interaction. Our novel approach, validated through various cooperative tasks, demonstrates enhanced performance and efficiency over traditional MARL baselines. Moreover, our results reveal the potential of ToCM to generalize and adapt rapidly to new tasks, showcasing its versatility and robustness in the domain of social intelligence.
\end{IEEEImpStatement}

\begin{IEEEkeywords}
Theory of collective mind model, multi-agent reinforcement learning, free-energy principle.
\end{IEEEkeywords}

\section{Introduction}

\IEEEPARstart{T}he internal or generative model in the brain used to infer one's own behaviour can be deployed to infer the beliefs of another - provided both parties have sufficiently similar generative models~\cite{ito2008control,haswell2009representation,friston2015duet}. This ability to infer the mental states of others is known as Theory of Mind (ToM), which allows people to imagine the process of interacting with others and learn from this contemplation. In multi-agent scenarios, the Theory of Collective Mind (ToCM) empowers an individual's mental representation of a unified, shared but plural mental perspective to a self-inclusive group of agents. Such a collective mind strengthens relational bonds and facilitates  cooperation~\cite{shteynberg2023theory}. ToCM is a natural and powerful approach to multi-agent decision making problems with incomplete information. 

With the development of Multi-Agent Reinforcement Learning (MARL), many related applications have emerged, such as real-time multi-agent games~\cite{berner2019dota}, autonomous driving~\cite{zhou2020smarts}, and robotic manipulation~\cite{zhou2020smarts}. ToCM offers several benefits to MARL. First, ToCM can address the issue of a non-fixed environmental state transition function due to changes in other agents because it infers rich information about others. Moreover, based on the inference about others, ToCM provides an imaginary space in which agents can interact, thereby shortening the learning process for agents in simulated environments. Finally, ToCM can be independent of any specific task and have the ability to transfer across different tasks within the environment.

The agent in the collective with ToCM needs to infer mental or physical states about oneself and others from shared experience. In complex simulated scenarios, just like ToM, ToCM is also systematic and accurate enough for low-energy inferring others, which has been a long-standing challenge. First, previous ToM-related studies were committed to modeling the mental state of a single individual, and when faced with multiple agents, it is necessary to assign each agent with a ToM model. As agents' number increases and the environment becomes more complex, the scale and number of ToM networks will grow explosively. Furthermore, ToM can infer many characteristics of others; its function is not limited to merely predicting others' behaviors or observations, but should also encompass other mental or physical states. Ultimately, the systematic model must also have the ability to solve complex problems. These problems include the disconnection between theoretical models and practical modeling as the scenario becomes more complex, the inaccuracy of predictions caused by agents with changing policies, and the failure to accurately infer others' multiple possible futures.

To address these issues, we draw inspiration from ToCM and design a unified, plural and collective mental state model capable of predicting future observations of self and others. 
First, this paper is inspired by the free energy principle (FEP)~\cite{friston2015duet,friston2015active,gershman2019does}, minimizing the free-energy to derive the objective function of ToCM. Moreover, based on this objective function, we construct an encoder that abstracts mental states from collective observations and multiple decoders that predict the physical states of others from mental states~\cite{baker_bayesian_2011, de2013higher, de2013much, de2014theory, baker_rational_2017}. Historical information of the collective constitutes contextual knowledge that can be utilized as a supervisory signal to aid in learning the decoding process from mental states to observation~\cite{rabinowitz_machine_2018, nguyen2023memory}. Finally, this paper constructs long-term inferences about the collective by utilizing shared information and mental states obtained through encoding.

%As the behaviors and positions of others change over time, it becomes necessary to seek an appropriate method for storing these time-varying sequences, such as RNNs~\cite{elman1990finding, graves2012long, cho2014learning}, Transformer~\cite{vaswani2017attention} and RSSM~\cite{hafner2019learning}. Further, mental states determine the future behavioral choices of others. It is evident that this process can be modeled as the posterior~\cite{baker_bayesian_2011, de2013higher, de2013much, de2014theory, baker_rational_2017}. Based on the inference of others' mental states, we can further predict their future performance, including behaviors, positions, and so on. Historical information can be utilized as a supervisory signal to aid in learning the decoding process from mental states to observation~\cite{rabinowitz_machine_2018, nguyen2023memory}.

When agents can predict the future information of the collective, they can use the inferred information to create an imaginative space and optimize their policies through imagination. To validate the effectiveness of our model, we conducted experiments on both simple tasks (Cooperative navigation and Heterogeneous navigation) and more complex tasks (Starcraft, as well as diverse decision networks such as Deep Neural Networks (DNNs) and Spiking Neural Networks (SNNs). For Starcraft, we also verified the transferability of the brain-inspired ToCM model.

Our main contributions can be summarized as follows:

1) We developed a brain-inspired ToCM model to learn the collective mental state from observed information, which in turn predicts future observations of self and others, helping to enrich the understanding of the current state.

2) The information inferred by the brain-inspired ToCM model can form an imaginative space that can simulate the environmental interaction of multi-agent to assist policy training, improve the learning speed and the knowledge utilization.

3) We perform experiments on various multi-agent cooperative tasks, compared with multiple MARL baselines, the proposed model can improve the cooperative efficiency and performance, and exhibit flexible transferability.

\section{Related Works}

\subsection{Theory of Mind Model}

ToM models allow an ego agent to infer the behavior, observations, rewards, and goals of others. Traditional research on ToM is based on Bayes' theorem, and the Bayesian Theory of Mind (BToM) model performs well in predicting others' desires based on their behavioral trajectories~\cite{baker_bayesian_2011, baker_rational_2017}. Recursive ToM models use Bayes' theorem and a greedy strategy, with first-order mental state inference utilizing a confidence parameter to balance the first-order and zero-order ToM models for the final inference~\cite{de2013higher, de2013much, de2014theory}. ToMoP algorithms further improve dynamically tunable confidence~\cite{yang2018towards} and have been used to solve two-agent decision-making tasks, including games like rock-paper-scissors and toy two-player games in the gridworld. ToMAGA uses higher-order ToM to estimate its influence on others, thereby reducing the negative impact of its actions on others in Stag Hunt Games~\cite{nguyen2020theory}. 

Besides, Bayesian inference is usually performed through inverse reinforcement learning (IRL). As described by~\cite{jara2019theory}, ToM is achieved by inverting a model that simulates the actions of others with hypothesized beliefs and desires using an RL model. The Bayesian model establishes the theoretical foundation for ToM, but it is limited to simple tasks, such as the gridworld. Recently, another Bayesian model demonstrated that a Bayesian agent has the capability to trigger effective interventions that enhance Human-AI team performance which inspired us to collectively
model the mental states of teammates~\cite{westby2023collective}.

Furthermore, researchers have explored the use of neural networks to model ToM. ToMnet, for example, combines neural networks with meta-learning to infer the mental states of others~\cite{rabinowitz_machine_2018}. ToMnet-G extends the original ToMnet by incorporating Graph Neural Networks (GNN) and advances its predictive capabilities from 2D to 3D, aiming to predict the trajectory of an agent within a 3D scene~\cite{shu2021agent}. Trait-ToM argues that stable personality traits provide essential prior information influencing mental states, improving performance over ToMnet using 'quick weights' and excels in inferring actors' mental states and goals based on their historical and present behavior~\cite{chuang2020using}. The present work introduces ToMMY, which utilizes advanced neural memory mechanisms and hierarchical attention to effectively infer the mental states of collective~\cite{nguyen2023memory}. Overall, these efforts aim to separate the observer from the environment and address the problem similar to trajectory prediction. 

Existing ToM methods utilize a separate function or network for each observed agent to infer their mental states, but with an increasing number of agents and their interdependent behaviors, the computational burden grows significantly and it becomes difficult to handle complex tasks. Therefore, in this work, we aim to use a single network to process the observations of any number of agents and obtain a collective mental state for both self and other agents.

\subsection{Multi-agent Reinforement Learning with Theory of Mind Model}

Just as humans rely on ToM mechanisms to interact with each other, we posit that ToM should be regarded as similarly essential in the context of multi-agent tasks. By enabling agents to contextualize and attribute mental states to other agents, ToM can facilitate improved cooperation, competition, and communication between agents. ToM2C employs historical information as a form of supervised signal to train agents on predicting the observations and objectives of others. By utilizing these inferences, the agents are able to make more informed decisions~\cite{wang2021tom2c}. MIRL-ToM is a novel approach to multi-agent inverse reinforcement learning that uses ToM to estimate the posterior distribution of an agent's reward profiles based on the observed behavior~\cite{wu2023multiagent}. SymmToM demonstrates that reinforcement learning agents modeling the mental states of others perform better~\cite{sclar2022symmetric}.
Additionally, a new explainable AI framework based on ToM has been developed to explain decision-making by deep convolutional neural networks~\cite{akula2019x, akula2022cx}.

Besides, opponent modeling has been utilized as a means of constructing an abstract model of other agents, serving as a proxy for theory of mind. This approach involves inferring the behaviors and rewards of opponents by explicitly representing their parameters~\cite{uther1997adversarial, ganzfried2011game, raileanu2018modeling} or implicitly extracting features~\cite{bard2013online, he2016opponent}, thereby optimizing one's own behavior. Even ToMoP draws inspiration from advanced theory of mind mechanisms, enabling multi-agents to collaborate more effectively in tasks.

Modeling the ToM in multi-agent scenarios remains a significant challenge and has its limitations. The challenges arise because the predictions of the ToM model cannot match the real mental state of other agents. The limitations lie in the inability to further complicate the experimental scenarios. This motivates us to explore the role of ToCM in complex cooperative scenarios with multiple agents. 

\section{Preliminaries and Notations}

Multi-agent reinforcement learning (MARL) is an extension of Markov decision processes (MDPs) to encompass scenarios with multiple agents. In a Markov game with $N$ agents, the system is characterized by hidden states $S={S_1,\cdots, S_N}$ (At time t, the hidden state of the collective can be represented as $ s_t $), actions $A={A_1,\cdots, A_N}$, and observations $O={O_1,\cdots, O_N}$. Each agent follows a policy to select actions, and generates the next state using a state transition function. Moreover, rewards received by each agent is: $R={R_1,\cdots, R_N}$. Following each agent's interaction with the environment, a trajectory ($o_1$, $a_{1}, r_{1}, o_{2},\cdots$) is generated and stored in buffer $\mathcal{D}$. We consider a setting with $N$ agents, where the agents are represented by policies (actors) $\pi = {\pi_1,\cdots, \pi_N }$ parameterized by $\theta={\theta_1,\cdots, \theta_N}$, and Q-functions (critics) $ Q^{\pi} = {Q^{\pi}_{1},\cdots, Q^{\pi}_{N} }$ parameterized by $\gamma={\gamma_1,\cdots, \gamma_N}$. N agents form a collective and jointly maintain a brain-inspired ToCM model, $ m_{\phi} $.

\section{Method}
In this article, we introduce an internal model called ToCM which enables mutual inference and prediction among agents within a collective by sharing information. This internal model creates an imaginary space to assist in the training of the decision-making model for the agents.

\subsection{Theory of Collective Mind Model (ToCM)}
\subsubsection{Establishing ToCM through Minimizing Free-energy.}
The FEP suggests that biological systems (like the brain) act to minimize their free-energy, which is thought to be related to minimizing surprise or prediction error over time~\cite{friston2015duet}.
At time $ t $, the agents perceive the external environment and receive observations $ o_t $, which include descriptions of the positions of other agents or objects, as well as velocity information for different agents. The observations serve as evidence for training ToCM. With evidence accumulation, ToCM gradually develops accurate inferences about the collective interaction process, including inferences about others' rewards, observations, and actions. Our objective is to minimize the average uncertainty about observations the agents receive. Minimizing free-energy Equation (\ref{fep}) can reduce uncertainty. The negative of log-evidence, $ -\ln (p(o)) $, is proportional to the uncertainty. $ p(o) $ is intractable because of the incomplete observations. To address the above issue, ToCM generated a variational distribution, $ q(s) $, to approximate $ p(s \mid o) $ where $ s $ is hidden state. This function is a Kullback–Leibler (KL) divergence $ D_{KL} $ and is greater than zero, with equality when $ q(s) = p(s \mid o) $ is the true conditional density. This means that minimizing free-energy, by changing $ q(s) $, makes ToCM an approximate conditional density on observations. 
\begin{equation}
	F=-\ln (p(o))+D_{KL}( q(s) \|   p(s \mid o))
	\label{fep}
\end{equation}

By the derivation of the formula and removing the intractable $ p(o) $, the free-energy has been expressed in terms of the cross entropy, $  -E_q(\ln p(o \mid s)) $, and KL divergence, $ D_{KL} $, in Equation (\ref{min}).

\begin{equation}
	 F =  -E_q(\ln p(o \mid s)) + D_{K L}( q(s) \|  p(s))
	\label{min}
\end{equation}

Based on the free-energy principle, we constructed the objective function for ToCM at time $ t $ in Equation (\ref{obj}). When ToCM infers collective mental states, it is influenced by the actions, $ o_t $, of each agent. Therefore, when calculating the KL divergence, we take actions into account.

\begin{equation}
	\begin{split}
	J&=-E_q\left(\ln p\left(a_t \mid s_t\right)\right)\\
	&-E_q\left(\ln p\left(o_t \mid s_t\right)\right)\\
	&-E_q\left(\ln p\left(r_t \mid s_t\right)\right) \\
	&+D_{K L}\left(q\left(s_t \mid s_{t-1}, a_{t-1}, o_t\right) \| p\left(s_t \mid s_{t-1}, a_{t-1}\right)\right)
	\end{split}
	\label{obj}	
\end{equation}

Based on the above rationale, we constructs ToCM, which consists of mental states encoder ($q\left(s_t \mid s_{t-1}, a_{t-1}, o_t\right)$), predictor, and mental states decoder modules. 
Moreover, we need to find three additional decoder networks to predict actions, observations, and rewards based on data acquired from the external environment. Finally, we need to establish a network that simulates the state transition equation.

\begin{figure}[H]
	
	\begin{center}
		\includegraphics[width=7.5cm]{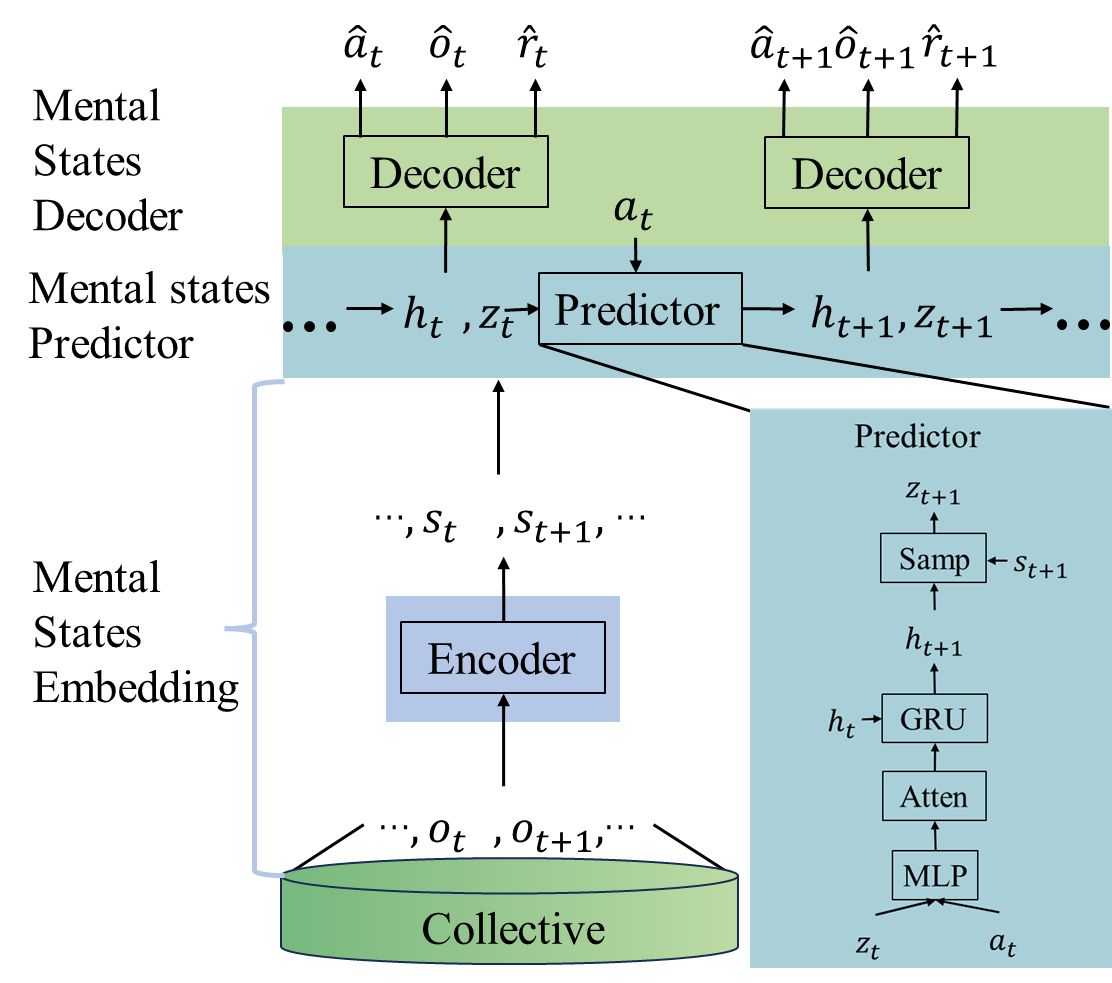}
	\end{center}
	\caption{Architecture of the brain-inspired ToCM model.}
	\label{tom}
\end{figure}

Fig. \ref{tom} illustrates the architecture of ToCM. Past observations involving information about collective are encoded into mental states, $ s_t $. The Predictor learns the relationship between consecutive hidden states (containing deterministic hidden states, $ h_t $ and stochastic hidden states, $ z_t $) over a long-term sequence by using RSSM~\cite{hafner2019learning}. Stochastic hidden states increase the model's uncertainty and exploration noise, while deterministic hidden states enhance the model's memory capacity. Based on the prediction of the next hidden states, the decoder outputs predictions about observations. 

\subsubsection{Mental States Embedding}
%\label{M1}
Mental states are a crucial concept in ToM, as they refer to the capacity to infer deeper abstract content from the representation of others. Inspired by this,
we conducted two bottom-up abstractions of the observations through the posterior, resulting in \textbf{metal states}, $s_t$ and another metal states, called \textbf{hidden states}. First, we have modeled the representation of mental states of others through the observation based on $ p\left(s_t \mid o_t\right) $. The observations of each agent is a one-dimensional vector that contains information about self as well as observations of others. The observation at time t is abstracted as metal states $ s_t $. The output mental states are also a one-dimensional vector. Furthermore, the mental states embedding module generates abstract hidden states (deterministic hidden states, $ h_t $ and stochastic hidden states, $ z_t $ ), based on $ s_t $ and $ a_{t-1} $. The deterministic hidden states are determined by the hidden state of the previous moment and the collective actions. The stochastic hidden states are derived from the current observation and mental states in Equation (\ref{rep}).
\begin{equation}
	s_t= \operatorname{MLP}\left( o_t\right)
	\label{post_network}
\end{equation}

\begin{equation}
	z_t=\operatorname{MLP}\left(h_{t}, s_t\right)
	\label{rep}
\end{equation}

Compared to assigning a separated posterior network to each agent, this approach has the advantage of a fixed number of networks, which does not increase with the number of agents, thus reducing the computational burden.

After attributing mental states to the collective, the agent with ToCM predicts the future mental states performances and rewards of others combining the actions.

\subsubsection{Predictor\label{M2}}
 The hidden states predictor recurrently produces abstract deterministic hidden states, $ h_{t+1} $ and stochastic hidden states, $ z_{t+1} $ based on $ h_{t} $, $ z_{t} $ and $ a_{t}$ by computing the prior, $p\left(h_{t}, z_{t} \mid h_{t}, z_{t}, a_{t}\right)$. 
 In the bottom-right diagram of Fig. \ref{tom}, the predictor module captures the relationship between adjacent hidden states through fully-connected layers, attention mechanisms and GRU~\cite{chung2014empirical}, and predicts the hidden states of the next moment by sampling.
 %Specifically, the hidden states $ h_{t} $ and $ z_{t} $ concatenated with $a_t$ are encoded as the input of a multi-layer perceptron combined with an attention mechanism to extract sequence features, and then use the output as input to a recurrent neural network for predicting the hidden state at the next time step.

\begin{equation}
	h_{t+1}, z_{t+1}= \operatorname{GRU}\left(\operatorname{Atten}\left(\operatorname{MLP}\left(h_{t}, z_{t}, s_{t+1}, a_t\right)\right)\right)
	\label{prior_network}
\end{equation}

\subsubsection{Mental States Decoder}  The next time step's hidden state $ h_{t+1} $ and $ z_{t+1} $ can be decoded into the observation of the next time step $ \hat{o}_{t+1} $ ($ o_{t+1} \sim p\left(o_{t+1} \mid h_{t+1}, z_{t+1}\right) $) by using MLP in Equation (\ref{decoder_network}). In order to enable the model to accurately infer the collective, the data generated by the decoder should be as close as possible to the collected real data. Rewards and behaviors are decoded from hidden states in the same manner.

\begin{equation}
	\hat{o}_{t+1}= \operatorname{MLP}\left(h_{t+1}, z_{t+1}\right)
	\label{decoder_network}
\end{equation}

\subsubsection{Loss Function:} Based on the
objective function (Equation (\ref{obj})), ToCM is trained by minimizing a action loss ($ L_{act} $), a observation loss ($ L_{obs} $),  a reward loss ($ L_{rew} $) and KL divergence ($ L_{\mathrm{KL}} $) :

\begin{equation}
	L = L_{act} + L_{obs} + L_{rew} + L_{\mathrm{KL}}
	\label{loss}
\end{equation}

We use the historical observations as labels to train the model for predicting others next positions based on the Huber loss~\cite{huber1992robust}. The reward loss and action loss are also computed using the Huber loss.

\begin{equation}
	L_{obs}(o_t, \hat{o}_t)= \begin{cases}\frac{1}{2}(o_t - \hat{o}_t)^2 & \text { for }|o_t - \hat{o}_t| \leq 1, \\ |o_t - \hat{o}_t|-\frac{1}{2}  & \text { otherwise. }\end{cases}
	\label{Huber}
\end{equation}

$ p\left(h_{t+1}, z_{t+1}  \mid h_{t}, z_{t}, a_t\right)  $ is the hidden state transition function and $ q\left(h_{t+1}, z_{t+1} \mid h_{t}, z_{t}, o_t\right) $ represents the inference of latent variables based on observations. Both of them are descriptions of latent variables. Therefore, we calculate the KL divergence between the two distributions:

\begin{equation}
	L_{\mathrm{KL}} = \mathrm{KL}[q\left(h_{t+1}, z_{t+1} \mid h_{t}, z_{t}, o_t\right) \| p\left(h_{t+1}, z_{t+1}  \mid h_{t}, z_{t}, a_t\right) ]
	\label{KL}
\end{equation}

\subsection{MARL with ToCM}
In this subsection, we describe how predictions obtained through ToCM help provide richer representations of the agent's current state for MARL, and 
offer an imaginative space where agents simulate environmental interactions with other agents.

\subsubsection{Interactions in Imagination}
\begin{figure}[H]
	
	\begin{center}
		\includegraphics[width=6cm]{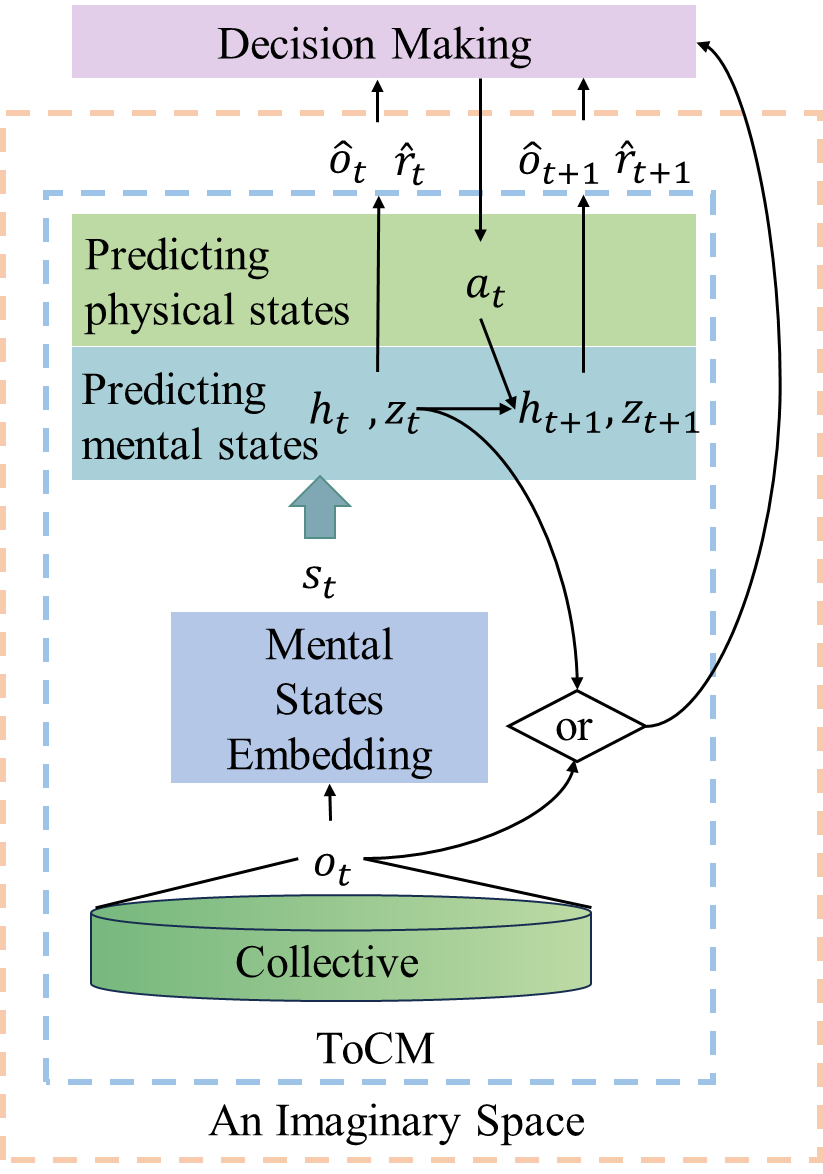}
	\end{center}
	\caption{Interaction between the decision-making model and a ToCM-based imaginative space. }
	\label{dec}
\end{figure}

By utilizing the brain-inspired ToCM model's ability to predict collective observations, rewards, and other information, we integrate this information into an imaginary space~\cite{hafner2020mastering, egorov2022scalable}, allowing the decision-making model to directly interact and learn within this imaginary space as shown in Fig. \ref{dec}. 
The collective with ToCM inferred the collective mental states $s_{t}$ and the observation of the next step $\hat{o}_{t+1}$. The collective makes a decision based on both the current information (the current observation $o_t$ or the hidden states $h_{t} $ and $ z_t $ ) and the predictive observation $\hat{o}_{t+1}$  and reward$\hat{r}_{t+1}$  of the next step. The predicted rewards make it possible for agents to learn policies in an imaginative space.

\begin{algorithm}[tb]
	\caption{MARL with ToCM}
	\label{alg:algorithm}
	\begin{algorithmic}[1] %[1] enables line numbers
		\STATE \textbf{for} t $ =1 $  to  epochs \textbf{do}   
		\STATE \quad \textbf{for} step $ =1 $  to  episode \textbf{do}  
		\STATE \quad  \quad For each agent  $ i  $, select action.  
		\STATE \quad \quad  Store $ \left(o, a, r, o^{\prime}\right) $ in replay buffer  $ \mathcal{D} $. 
            \STATE \quad \textbf{ end for }
            \STATE \quad   \# Train ToCM
		\STATE \quad \textbf{for} $t_m$ $ =1 $  to  model\_epochs \textbf{do}              
		\STATE \quad  \quad Inferring the mental states from Eq. \ref{post_network}.
		\STATE \quad \quad  Calculating the hidden state from Eq. \ref{rep}.
		\STATE \quad  \quad Predict actions, observations and rewards \\ based on mental states decoder.
		\STATE \quad \quad  Train ToCM with Eq. \ref{loss}.
            \STATE \quad \textbf{ end for }
  		\STATE  \quad  \# Train MARL
            \STATE \quad \textbf{for} $t_{RL}$  $=1 $  to  MARL\_epochs \textbf{do}   
		\STATE \quad \quad  Train policy $\pi_\theta$ in the imaginative space generated from ToCM for max-episode length.
		\STATE  \quad \quad Compute advantage estimates $ Ad $.
		\STATE  \quad \quad Optimize surrogate $\mathcal{L}_{actor}$ with Eq. \ref{actor}.
		\STATE \quad  \quad $\theta_{\text {old }} \leftarrow \theta$.
            \STATE \quad \textbf{ end for }
		\STATE \textbf{ end for }
	\end{algorithmic}
\end{algorithm}
\subsubsection{MARL Model} 

We chose independent proximal policy optimization (IPPO)~\cite{schulman2017proximal, rashid2018qmix, de2020independent} as decision making method. Following IPPO, actor and critc are update by minimizing the actor loss in Eq. \ref{actor}, critic loss $ \left(V_\gamma\left(o_t\right)-V_t^{\operatorname{targ}}\right)^2 $ and an entropy bonus of policy $\pi$:
\begin{equation}
	\begin{aligned}
		&\mathcal{L}_{actor}(\theta)=\mathrm{E}_{o_t, a_t}\left[\min \left(\frac{\pi_\theta\left(a_t \mid o_t, \hat{o}_{t+1}\right)}{\pi_{\theta_{\text {old }}}\left(a_t \mid o_t, \hat{o}_{t+1}\right)} Ad\left(o_t, a_t\right), \right.\right.\\
		&\left.\left.\operatorname{clip}\left(\frac{\pi_\theta\left(a_t \mid o_t, \hat{o}_{t+1}\right)}{\pi_{\theta_{\text {old }}}\left(a_t \mid o_t, \hat{o}_{t+1}\right)}, 
		1-\epsilon, 1+\epsilon\right) Ad\left(o_t, a_t\right)\right)\right]
	\end{aligned}
	\label{actor}
\end{equation}
$ \pi $ represents the policy function, $ V $ represents the value function, $ Ad $ represents the advantage function and $ \epsilon $ a hyperparameter. The advantage function measures the relative advantage of choosing an action in a given state compared to the average action. The value function measures the expected return in a given state. The policy function defines the probability distribution over actions to take in a given state. These functions are commonly used to describe value estimation and decision-making processes in reinforcement learning.

In this paper, we mainly consider the SNN-based MARL model because SNNs, as the third-generation neural networks, show more potential for future multi-robot applications due to their high energy efficiency and suitability for execution on neuromorphic processors. In addition, to verify the generalizability of our model, we conducted experiments on both SNN- and DNN-based decision-making network. For the SNN-based actor network, we adopted the leaky integrate-and-fire (LIF) neuron~\cite{b26, b33} and trained it with the surrogate gradient~\cite{wu2018spatio}.

\section{Experiment Results}
\subsection{Environments}
\begin{figure}[H]
	\begin{center}
		\includegraphics[width=7cm]{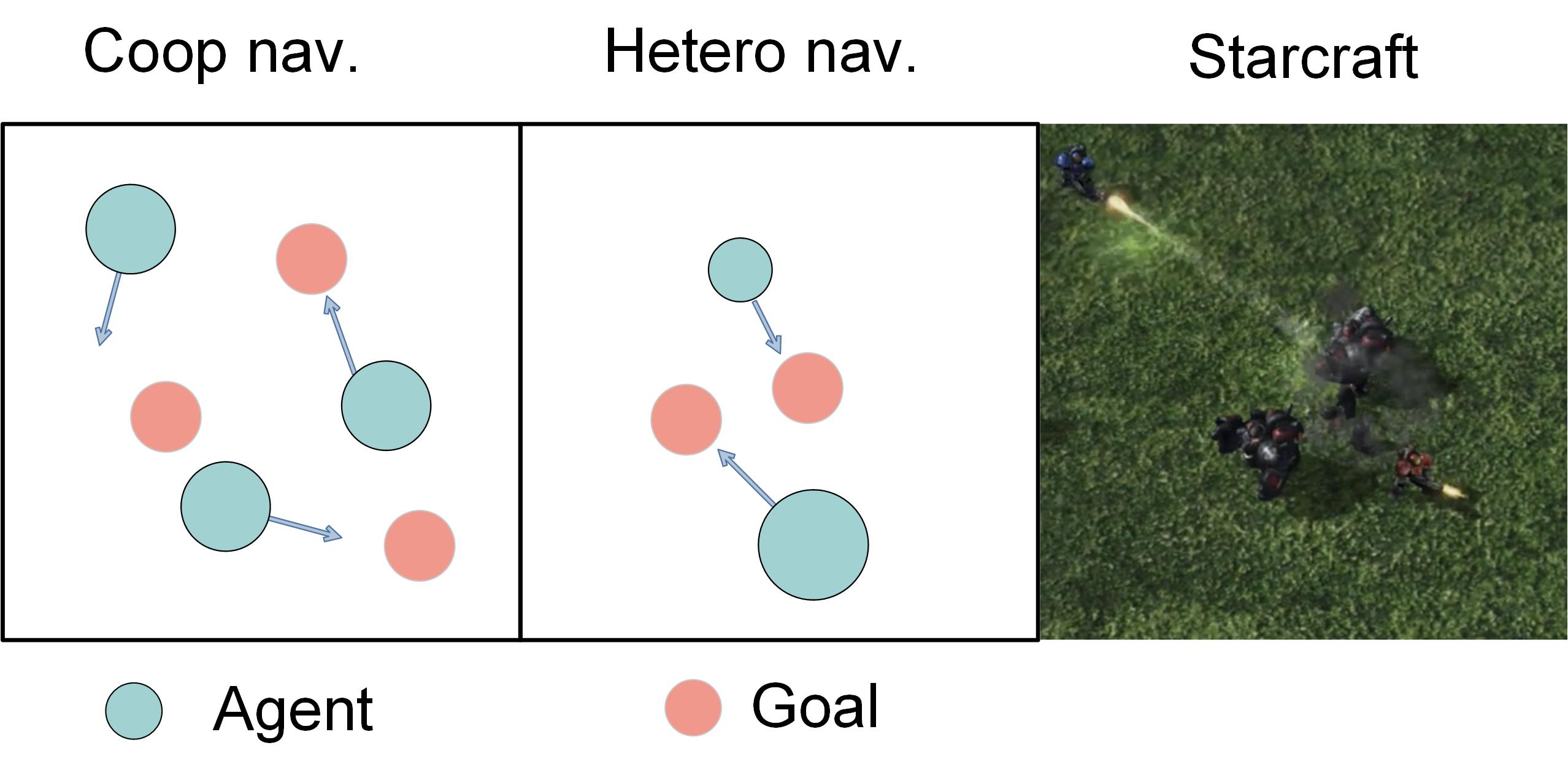}
	\end{center}
	\caption{Illustrations of the multi-agent cooperative environments. Left: Cooperative navigation. Middle: Heterogeneous navigation. Right: SMAC Multi-Agent Challenge~\cite{samvelyan19smac}}
	\label{star}
\end{figure}

We evaluate the ToCM model on multiple cooperative tasks, including the multi-agent particle environments (MPEs)~\cite{lowe2017multi, mordatch2017emergence} and SMAC Multi-Agent Challenge (SMAC)~\cite{samvelyan19smac} as shown in Fig. \ref{star}. MPEs involving cooperative navigation and heterogeneous navigation are fully observable tasks. SMAC is a partially observable task.

\textbf{Cooperative navigation.} To achieve a group of $N$ goals, $N$ agents need to work together. The agents will receive a shared reward based on the presence of any agent at any goal location. At each step, each agent's action is chosen  between five primitive actions $\{no\_action, move\_left, move\_right, move\_down,\\ move\_up\}$. To quickly cover all the goals, each agent needs to infer the goals of others to avoid heading towards the same target as them. Each episode has 25 steps. In the experiment, $N$ is set to 2, 3, and 4. 

\textbf{Heterogeneous navigation.} This task is similar to cooperative navigation, the difference is: there are $N$ goals that need to be reached by a group of $N$ agents, but they vary in size and speed. Specifically, there are two groups of agents: $ \frac{N}{2} $  slow and large agents, and $ \frac{N}{2} $ fast and small agents. In this experiment, $N$ is set to 2, and 4. 

\textbf{SMAC.} The SMAC is a set of cooperative multi-agent environments that are based on StarCraft II. Each task involves a single scenario with two opposing teams, one of which is controlled by the game bot and the other by our algorithm. We selected four scenarios for experimentation. These four scenarios are $  2s\_vs\_1sc $ (2 Stalkers against 1 Spine Crawler, the episode length is 300), $ 3s\_vs\_3z $ (3 Stalkers against 5 Zealots, the episode length is 150), $ 3m$ (3 Marines against 3 Marines, the episode length is 120) and $2s3z$ (2 Stalkers \& 3 Zealots against 2 Stalkers \& 3 Zealots, the episode length is 120), respectively. The game concludes either after 300 steps or once a victor is determined between the two sides
An episode ends when a victor is determined between the two sides or the number of steps exceeds the limit.

\subsection{Training and Evaluation}
As shown in the Fig. \ref{dec}, multiple agents share a unified ToCM and can invoke it by feeding their own observations. The ToCM encodes a collective mental states based on current observations, and predicts the future information of self and others. Furthermore, combined with the prediction of rewards, the decision-making model can be trained in an imaginative space, which greatly reduces the number of interactions between the decision-making model and the environment.
The advantage of playing in imagination is that agents can effectively utilize information in the environment, enhancing sample efficiency. For MPEs, we train the ToCM and MARL together with 3 seeds. However, for SMAC, we found that building an appropriate imaginative space while simultaneously learning policies is inefficient. This is because setting up an imaginative space for a task with intricate interpersonal relationships requires consuming a substantial amount of samples. Therefore, we pretrain ToCM for 200K timesteps iterations and employ the pretrained parameters for agents executing various tasks in SMAC. Similarly, each task in SMAC was subjected to three random experiments.  For comparison, we implemented value-decomposition networks (VDN)~\cite{sunehag_value-decomposition_2017}, MADDPG~\cite{lowe2017multi}, QMIX~\cite{rashid2020monotonic} and MAPPO~\cite{yu2022surprising} as baselines.

\subsection{Observation Prediction}

\begin{figure}[H]
	
	\begin{center}
		\includegraphics[width=8cm]{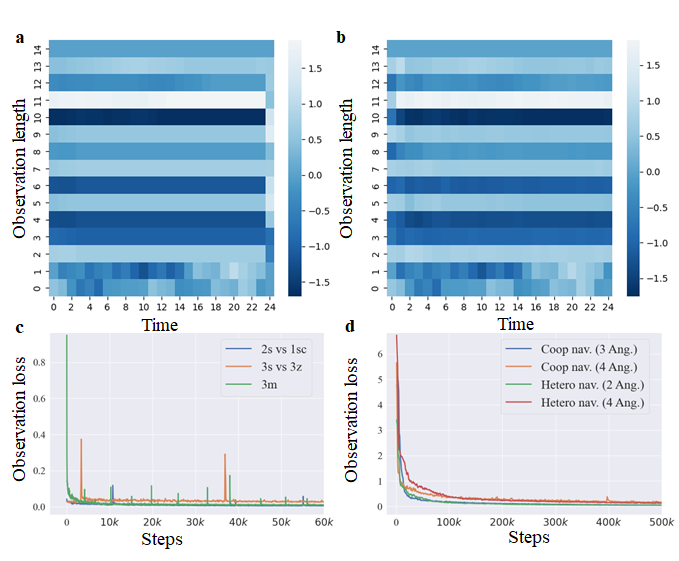}
	\end{center}
	
	\caption{The results of observation prediction. The comparison between the real observations (\textbf{a}) with a one-dimensional vector and the predictive observations (\textbf{b}). The transition from light to dark colors indicates a change in the magnitude of the observation values. \textbf{c}, \textbf{d}: The observation loss of ToCM across multiple tasks.}
	\label{heat1}
\end{figure}

Accurately predicting observations is crucial, as it lays the foundation for constructing an imagination space, which in turn assists in training policies.
Although the mental state is an unobservable variable, since observations are decoded from the collective mental states, we use observations as a metric to evaluate ToCM. We present predictions of the observations for the i-th agent.
We use the hidden state $h_t^{i}$ and $z_t^{i}$ combined with actions to predict the hidden state of the next timestep $h_t^{i}$ and $z_t^{i}$, and subsequently predict the observations for that next timestep $\hat{o}_{t+1}^{i}$.
We take cooperative navigation with 3 agents where the observation is one-dimensional vector with continuous values as an example to demonstrate: we recorded the predicted values of $\hat{o}_{t}^{i}$ over 25 steps ($t=0,\dots,24$) as shown in Fig. \ref{heat1}. The observation $o_{t}^{i}$ with the length of 15 involved
in self $o_{t}^{i}[0:4]$ and others $o_{t}^{i}[-5:-1]$ observation information.

From Fig. \ref{heat1}\textbf{a}, \textbf{b}, we can find that for the predicted observations of each agent (y-axis), at different time steps (x-axis), as well as for the value of observations (color), our ToCM model is able to achieve a level close to the real one, which indicates the accuracy of our model in predicting the observations of self and others, and thus verifies the ability to predict the collective mental state. Besides, we presented the loss of the real observation and the predictive observation on both MPEs and SMAC tasks as shown in Fig. \ref{heat1} \textbf{c}, \textbf{d}. 
The continuous convergence of the loss demonstrates that ToCM performs well on each task. 

All results verify that ToCM possesses the capability to accurately predict long-term future observations  based on current observations. This lays the foundation for constructing an imagination space, enabling agents to efficiently utilize environmental information to accelerate policy learning.

\subsection{Effects of Transferring Mental States}
\begin{figure*}[h]
	\centering
	\includegraphics[width=14.1cm]{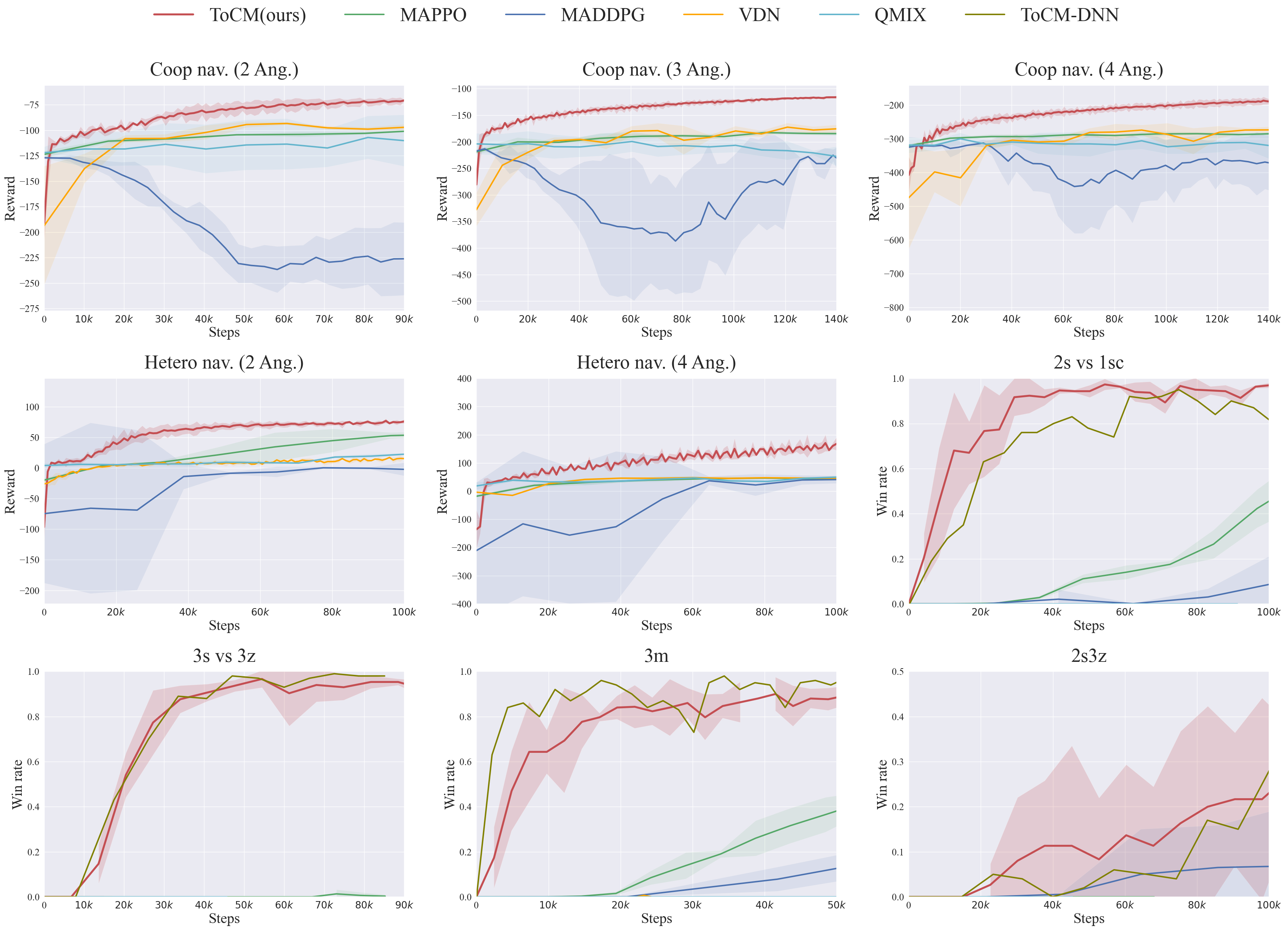}
	\caption{Comparative results of different algorithms in the MPEs and SMAC environments. Y axis represents episode rewards in MPEs and winning rate in SMAC, while X axis denotes number of steps taken in the environment. 
	}
	\label{re1}
\end{figure*}
In contrast to traditional methods that establish a ToM network for each agent when inferring others, we employ a single ToCM model to represent collective mental states. This approach reduces the number of networks and the amount of parameters. Additionally, to make the computation more concise, we consider drawing inspiration from the way that the human brain continuously deals with different tasks: leveraging the capabilities learned in scenario 1 to handle tasks in scenario 2. Therefore, we attempt to accelerate the learning speed of agents in new environments by employing the transferring mental states (TMS) achieved by ToCM.
\begin{figure}[H]
	
	\begin{center}
		\includegraphics[width=9cm]{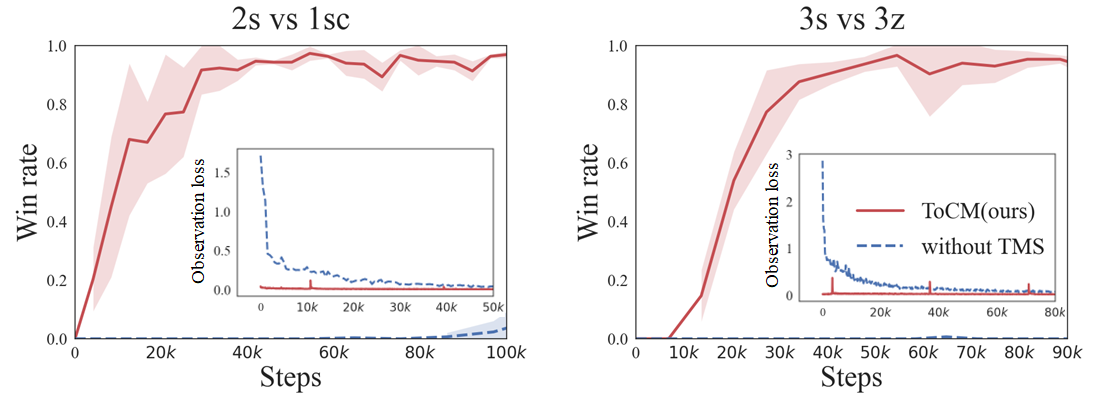}
	\end{center}
	
	\caption{Comparison of convergence speed with and without transferring mental states. Y axis represents winning rate in SMAC, while X axis denotes number of steps taken in the environment.}
	\label{spead}
\end{figure}

We found that TMS is more beneficial for complex tasks and thus conducted analysis in the SMAC environment. Inspired by children's learning of multiple tasks from simple to complex, we transferred the mental states learned in one scenario to another tasks. Specifically, we conducted a pretraining for 200K timesteps on the $2s3z$ task and used the trained weights as prior knowledge for the $  2s\_vs\_1sc $ and $  3s\_vs\_3z $ task.
Fig.\ref{spead} shows that the comparison with or without transferring the mental states. 
It can be found that TMS (the red lines in Fig.\ref{spead}) aids agents in rapidly adapting to new scenarios with minimal exploration.

To analyze the reasons for this advantage, we displayed the changes in ToCM's loss in new scenarios with and without transferring the mental states. The results show that TMS uses historical experience as prior knowledge for new tasks and results in a relatively small observation loss of ToCM in the initial timesteps, thereby reducing the environmental data needed by ToCM to model others and accelerating the learning of the policy network. A few spike-like fluctuations in the loss curves might be due to ToCM encountering features that are entirely different from previous experiences.

\subsection{Efficient Cooperation with ToCM}
We integrated ToCM into spiking actor based MARL and conducted experiments on MPEs and SMAC with different number of agents. For SMAC, we adopted the TMS to help agents quickly adapt to scenarios with complex role relationships. For simple tasks like MPEs, we directly adopt the ToCM without TMS. Specifically, we take the well-performing MARL algorithms involving MAPPO, MADDPG, VDN and QMIX as comparison. Since existing ToM models are mostly tested in specially designed non-generic small-scale grid environments and cannot be applied to complex multi-agent interaction tasks, we did not compare with the current ToM models. 

Fig. \ref{re1} displays results from three cooperative navigation tasks, two heterogeneous navigation tasks, and four SMAC tasks. For the cooperative navigation and heterogeneous navigation tasks, the score is an average reward of an agent in one episode. For the SMAC experiments, the result is based on the win rate. In MPEs, compared to the MARL baselines, our ToCM model (the red lines in Fig. \ref{re1}) remarkably improves the performance, accelerates the convergence speed and maintains high stability. When the ToCM curve begins to converge, the MARL baselines are still in the learning phase, causing fluctuations in the curve. Because the imagination space built by ToCM fully utilizes the sampling data in a short time to assist in policy training, the ToCM curve converges faster.

For more complex SMAC tasks, DNNs are needed to solve these complex tasks, and thus we also compared the results with non-spiking versions (ToCM-DNN) to verify the ToCM's generalizability. The experimental results show that our proposed ToCM-based models significantly outperform the MARL baselines when tackling complex tasks. The superior results based on SNN and DNN decision-making network also demonstrate the generalizability of the brain-inspired ToCM model.

\subsection{Comparative Experiments and Analysis}
In the preceding subsection, we juxtaposed the MARL with ToCM with widely used MARL algorithms, thereby illustrating the advantages of ToCM. In this subsection, we have conducted comparative experiments between ToCM and MAToM~\cite{zhao2023brain}, which makes agents predict other agents' actions and make decisions while considering the future actions of others in five tasks of MPEs. 
\begin{table}[ht]
\setlength{\tabcolsep}{4pt}

\centering
\caption{Comparison results between two ToM methods.}
\begin{tabular}{c|ccc}
\toprule
 Coop nav. & 2 Ang. &3 Ang. &4 Ang. \\
\midrule
MAToM~\cite{zhao2023brain} & $-118.39 \pm 0.09$ & $-203.10 \pm 0.30$ & $-299.30 \pm 0.21$ \\
ToCM(ours) & $-85.61 \pm 2.26$ & $-137.64 \pm 2.51$ & $-219.51 \pm 6.55$ \\

% Row 3, Col 1 & Row 3, Col 2 & Row 3, Col 3 & Row 3, Col 4 \\
% \bottomrule
% \toprule
\midrule
Hetero nav. & 2 Ang. & 4 Ang. \\
\midrule
MAToM~\cite{zhao2023brain} & $4.84 \pm 0.04$ & $36.81 \pm 0.29$  \\
ToCM(ours) & $57.22 \pm 3.18$ & $108.98 \pm 6.77$ \\

% Row 3, Col 1 & Row 3, Col 2 & Row 3, Col 3 & Row 3, Col 4 \\
\bottomrule
\end{tabular}
\label{comp}
\footnotesize\centering{The table shows the mean and standard deviation of episode rewards.} 
\end{table}
\medskip % Adds a medium vertical space

The primary rationale behind selecting MAToM as the baseline lies in its novelty and task complexity. MAToM represents a contemporary methodology that not only draws upon existing work for modeling but also extends its applicability to multi-agent interactive tasks.
The results are presented in Table \ref{comp} which recorded the mean and standard deviation of episode rewards for training 300K steps. In five experimental tasks, ToCM demonstrates a substantial advantage over MAToM within a short-term training period. Nonetheless, one minor concern arises: ToCM exhibits a higher standard deviation. Due to the limited data available in the early stages of training, the brain-inspired ToCM model was not adequately trained, resulting in poor initial performance. However, the overall performance of ToCM significantly surpasses that of MAToM.

\section{Conclusion}
To enable efficient cooperation of multiple agents in complex scenarios, we introduced ToCM, an efficient collective ToM model that predicts self and others into a unified and plural mental states representation. ToCM aids decision making in two ways: the predicted collective observations serve as inputs for decision making, enriching the understanding of the current environment; and a imaginative space is constructed based on ToCM accelerates the learning of the policy network. The advantage of collective mental states is that the number of network parameters does not increase with the number of modeled agents. This effectively mitigates the impact of the number of agents on the size of network parameters, saving computational resources. Furthermore, the imagination space can help agents simulate real-world interactions, enhancing sample efficiency. Extensive experiments on cooperative tasks validate the effectiveness of the proposed model, achieving higher performance and faster learning. Besides, our algorithm exhibits transferability and generalization to some extent. Overall, this paper establishes a brain-inspired ToCM model that addresses complex multi-agent cooperative tasks. In the future, we will further delve into constructing ToCM models to apply to specific human-computer interaction scenarios, and attempt to guide AI to develop moral intuition through ToM and other social cognition functions.
%\section*{References}
\bibliography{Abib}

\end{document}